\documentclass[twocolumn,pra,groupedaddress]{revtex4}

\usepackage{graphicx}
\usepackage{dcolumn}
\usepackage{bm}
\usepackage{amssymb}
\usepackage{amsfonts}
\usepackage{subfig}
\usepackage{latexsym}
\usepackage{enumerate}
\usepackage[dvipdfmx]{color} 
\usepackage{amsmath}
\usepackage[dvipdfmx]{epsfig}
\usepackage{times}
\usepackage{caption}

\newtheorem{theorem}{Theorem}
\newtheorem{lemma}{Lemma}
\newtheorem{definition}{Definition}
\newtheorem{conjecture}{Conjecture}

\begin{document}
\widetext
\title{Ancilla-driven instantaneous quantum polynomial time circuit for quantum supremacy}
\author{Yuki Takeuchi${}^1$ and Yasuhiro Takahashi${}^2$}
\affiliation{$^1$Graduate School of Engineering Science, Osaka University, Toyonaka, Osaka 560-8531, Japan\\
$^2$NTT Communication Science Laboratories, NTT Corporation, Atsugi, Kanagawa 243-0198, Japan}

\begin{abstract}
Instantaneous quantum polynomial time (IQP) is a model of (probably) non-universal quantum computation. Since it has been proven that IQP circuits are unlikely to be simulated classically up to a multiplicative error and an error in the $l_1$ norm, IQP is considered as one of the promising classes that demonstrates quantum supremacy. Although IQP circuits can be realized more easily than a universal quantum computer, demonstrating quantum supremacy is still difficult. It is therefore desired to find subclasses of IQP that are easy to implement. In this paper, by imposing some restrictions on IQP, we propose ancilla-driven IQP (ADIQP) as the subclass of commuting quantum computation suitable for many experimental settings. We show that even though ADIQP circuits are strictly weaker than IQP circuits in a sense, they are also hard to simulate classically up to a multiplicative error and an error in the $l_1$ norm. Moreover, the properties of ADIQP make it easy to investigate the verifiability of ADIQP circuits and the difficulties in realizing ADIQP circuits.
\end{abstract}

\maketitle 
\begin{center}
\noindent{\bf I. INTRODUCTION}
\end{center}
A universal quantum computer can solve some problems that seem to be intractable for a classical computer, such as integer factorization~\cite{[S97]}. However, the rigorous relationship between classical and quantum computers has not yet been specified. Recently, in order to develop an understanding of the difference between them, the classical simulatability of restricted quantum computation has been extensively studied. For such quantum computation, commuting quantum computation including instantaneous quantum polynomial time (IQP)~\cite{[SB09],[BJS11],[NN13],[TTYT16],[BMS16],[GWD16]}, deterministic quantum computation with 1 pure qubit (DQC1)~\cite{[KL98],[MFF14],[FKMNTT15]}, boson sampling~\cite{[AA13],[LLKROR14],[B15],[HGPMG15]}, constant-depth quantum circuit~\cite{[TD04],[TYT14]}, and permutational quantum computing~\cite{[J10]} have been proposed.

It has been proven that IQP circuits, DQC1 circuits, and boson samplers are unlikely to be simulated classically up to a multiplicative error~\cite{[BJS11],[FKMNTT15],[KLM01]}. However, since a multiplicative error is unnatural, it is also desirable to show such unlikeliness in the case of an error in the $l_1$ norm. In 2013, Aaronson and Arkhipov showed this for boson sampling under two conjectures~\cite{[AA13]}. By generalizing their argument, Bremner {\it et al.} showed that IQP circuits are unlikely to be simulated classically up to an error in the $l_1$ norm whose value is constant under only one conjecture~\cite{[BMS16]}.

In order to demonstrate quantum supremacy by using IQP circuits, we have to generate and measure a complex entangled state. To do that, all of the qubits must have a long coherence time and high controllability. However, it is difficult to prepare qubits that satisfy these conditions. In fact, proof-of-principle experiments have already been implemented for DQC1~\cite{[LBAW08]} and boson sampling~\cite{[BFKDARW13],[SMHKJBDPLKGSSW13],[TDHNSW13],[CORBGSVMMS13],[BSVFVLMBGCROS15]}, but have not yet been implemented for IQP. In other words, IQP circuits have an elegant mathematical structure, but they are not easy to implement.

As a model of universal quantum computation, ancilla-driven quantum computation (ADQC)~\cite{[AOKBA10]} has been proposed, which is a model intermediate between the gate-based model~\cite{[NC00]} and measurement-based quantum computation (MBQC)~\cite{[RB01]}. Since ADQC is realized by a fixed interaction between a register qubit and an ancillary qubit and by single-qubit operations for an ancillary qubit, it is suitable for many experimental settings such as those in Refs.~\cite{[CZ00],[CDJWZ04]}.
 
In this paper, by imposing some restrictions on IQP, we propose ancilla-driven IQP (ADIQP) as a subclass of commuting quantum computation. In ADIQP, qubits can be divided into white qubits with long coherence time but without high controllability and black qubits with high controllability but without long coherence time. Since a fixed interaction is applied only between a white qubit and a black qubit, the generated graph state is a two-colorable graph state. From these properties, ADIQP is suitable for many experimental settings like ADQC is. We show that even though ADIQP circuits are strictly weaker than IQP circuits in a sense, they are also hard to simulate classically up to a multiplicative error and an error in the $l_1$ norm. As with IQP, we require only one conjecture to show the latter statement. In contrast, different from IQP, the value of an error in the $l_1$ norm depends on the size of the ADIQP circuit. However, by removing one restriction from ADIQP, we can prove it in the case of an error in the $l_1$ norm whose value is constant. Moreover, toward the experimental realization of ADIQP circuits, we consider their verification to certify that a generated graph state is the desired one and investigate the difficulties in realizing them. From this consideration, it is shown that realizing ADIQP circuits is worthwhile not just for demonstrating quantum supremacy but also for realizing a fault-tolerant universal quantum computer.

\medskip
\begin{center}
\noindent{\bf II. INSTANTANEOUS QUANTUM POLYNOMIAL TIME} 
\end{center}
To clarify the difference between IQP and ADIQP, we first review the IQP briefly.

\begin{figure}[t]
\begin{center}
\includegraphics[width=5cm, clip]{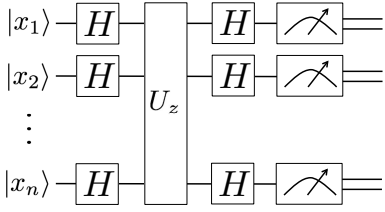}
\end{center}
\captionsetup{justification=raggedright,singlelinecheck=false}
\caption{The general IQP circuit. Each of the meter symbols represents the measurement in the $Z$ basis.}
\label{IQP}
\end{figure}

\begin{definition}
An IQP circuit (Fig.~\ref{IQP}) on $n$ qubits is defined as a quantum circuit that satisfies the following conditions:
\begin{itemize}
\item the input state is $|x\rangle\equiv|x_1\cdot\cdot\cdot x_n\rangle$ $(x_i\in\{0,1\},1\le i\le n)$,
\item the quantum gate is $H^{\otimes n}U_zH^{\otimes n}$, where $U_z$ is any $n$-qubit diagonal gate in the Pauli Z basis, and
\item the measurements are Pauli Z-basis measurements.
\end{itemize}
Here, $|0\rangle$ $(|1\rangle)$ is a $+1$ $(-1)$ eigenstate of the Pauli Z gate $Z$ and $H$ is a Hadamard gate. $H$ satisfies that $H|0\rangle=(|0\rangle+|1\rangle)/\sqrt{2}\equiv|+\rangle$ and $H|1\rangle=(|0\rangle-|1\rangle)/\sqrt{2}\equiv|-\rangle$.
\end{definition}

As an example of IQP circuits, we consider the case where $|x\rangle=|0\rangle^{\otimes n}$ and $U_z=\Pi_{i=1}^nZ_i(\theta_i)G_{b.w.}$. Here, $Z_i(\theta_i)\equiv e^{-i\theta_iZ_i/2}$, $Z_i$ is performed on the $i$th qubit, and $G_{b.w.}|+\rangle^{\otimes n}$ is a graph state called the $n$-qubit brickwork state~\cite{[BFK09]}. Since any $n$-qubit graph state is generated from $|+\rangle^{\otimes n}$ by using only the controlled-$Z$ gate $\Lambda(Z)\equiv|0\rangle\langle 0|\otimes I+|1\rangle\langle 1|\otimes Z$, $G_{b.w.}$ is the diagonal gate in the $Z$ basis. Here, $I$ is a two-dimensional identity operator. Since the brickwork state is a universal resource of MBQC~\cite{[BFK09]}, IQP with feedforward operations is equivalent to universal quantum computation.

We review classical simulatability of the IQP circuits in the case of a multiplicative error and an error in the $l_1$ norm. If there is a randomized classical circuit $R_n$ on $n$ bits with input $x$ that satisfies $|{\rm Pr}[R_n {\rm \ outputs\ } y {\rm \ on\ } x]-{\rm Pr}[C_n {\rm \ outputs\ } y {\rm \ on\ } x]|\le c{\rm Pr}[C_n {\rm \ outputs\ } y {\rm \ on\ }x]$ $(c\ge1)$ in polynomial time, we say that the output probability distribution of the quantum circuit $C_n$ on $n$ qubits is classically simulatable up to a multiplicative error $c$.
\begin{theorem}{\rm \cite{[BJS11]}}
If the output probability distribution of any IQP circuit is classically simulatable up to a multiplicative error $1\le c<\sqrt{2}$, then the polynomial hierarchy (PH) collapses at the third level.
\end{theorem}

PH is an infinite tower of complexity classes including non-deterministic polynomial time (NP)~\cite{[P94]}, and it is widely believed that PH does not collapse at any level.

Recently, a similar statement has been proven in the case of an error in the $l_1$ norm under one conjecture.
\begin{conjecture}{\rm \cite{[BMS16]}}
Let $f:\{0,1\}^n\rightarrow\{0,1\}$ be a uniformly random degree-$3$ polynomial over $\mathbb{F}_2$ {satisfying} $f(0^n)=0$. Then it is \#P-hard to approximate $[{\rm gap}(f)/2^n]^2$ up to a multiplicative error $c=1/4+o(1)$ for a $1/24$ fraction of polynomials $f$. Here, ${\rm gap}(f)\equiv|\{x:f(x)=0\}|-|\{x:f(x)=1\}|$.
\end{conjecture}
\begin{theorem}{\rm \cite{[BMS16]}}
Assume conjecture 1 is true. If the output probability distribution of any IQP circuit is classically simulatable up to an error of $1/192$ in the $l_1$ norm, then the PH collapses at the third level.
\end{theorem}

Here, \#P is a complexity class consisting of function problems that can be solved by counting the number of solutions of arbitrary NP problems~\cite{[P94]}.

\begin{figure}[t]
\begin{center}
\includegraphics[width=5cm, clip]{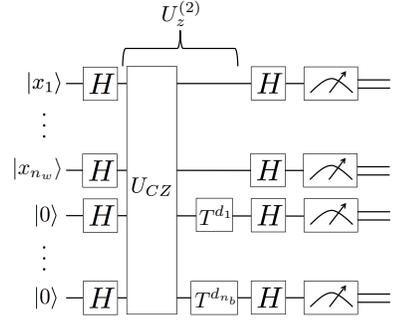}
\end{center}
\captionsetup{justification=raggedright,singlelinecheck=false}
\caption{The general ADIQP circuit. $U_{\rm CZ}$ is a quantum gate composed of $\Lambda(Z)$, and $U_z^{(2)}=\Pi_{l=1}^{n_b}T_l^{d_l}U_{\rm CZ}$ $(0\le d_l\le 7)$.}
\label{ADIQP}
\end{figure}

\medskip
\begin{center}
\noindent{\bf III. ANCILLA-DRIVEN INSTANTANEOUS QUANTUM POLYNOMIAL TIME}
\end{center}
By imposing some restrictions on IQP, we propose a subclass of commuting quantum computation suitable for many experimental settings.
\begin{definition}
An ADIQP circuit (Fig.~\ref{ADIQP}) on $n$ qubits is defined as a quantum circuit that satisfies the following conditions:
\begin{itemize}
\item the input state is the product of $|x\rangle\equiv|x_1\cdot\cdot\cdot x_{n_w}\rangle$ $(x_j\in\{0,1\}, 1\le j\le n_w)$ and $|0\rangle^{\otimes n_b}$ $(n_w+n_b=n)$, where the $n_w$ qubits in the state $|x\rangle$ and the $n_b$ qubits in the state $|0\rangle^{\otimes n_b}$ are called white qubits and black qubits, respectively, 
\item the quantum gate is $H^{\otimes n}U_z^{(2)}H^{\otimes n}$, where $U_z^{(2)}$ is any $n$-qubit diagonal gate in the $Z$ basis that is composed of $\Lambda(Z)$ between a white qubit and a black qubit and $Z(\pi/4)\equiv T$ on a black qubit, and only two or less $\Lambda(Z)$ can be performed on a black qubit,
\item the measurements are $Z$-basis measurements, and
\item the input register and output register are composed of white qubits.
\end{itemize}
Here, the superscript $(2)$ means that $U_z^{(2)}|+\rangle^{\otimes n}$ is a two-colorable graph state up to local unitary operations.
\end{definition}

Since a black qubit is measured in the basis lying on the equator of the Bloch sphere after it has interacted with at most two white qubits, a black qubit is not required to have long coherence time, but it has to be easy to manipulate. On the other hand, since a white qubit is measured in the Pauli $X$-basis after it has interacted with many black qubits, a white qubit is required to have long coherence time, but it does not have to be easy to manipulate. In other words, to realize ADIQP circuits, we can use different physical systems as white qubits and black qubits. Moreover, we do not have to realize the interaction between identical systems such as that between two photonic qubits. Note that some white qubits can be used as ancillary qubits, which are used to entangle two black qubits.

We consider the relationship between IQP and ADIQP. From definitions 1 and 2, it is obvious that IQP circuits can simulate ADIQP circuits exactly. We show that ADIQP circuits are strictly weaker than IQP circuits in the following sense.
\begin{figure}[t]
\begin{center}
\includegraphics[width=7cm, clip]{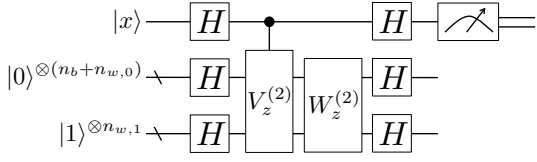}
\end{center}
\captionsetup{justification=raggedright,singlelinecheck=false}
\caption{The ADIQP circuit in the case where the input register is the same as the output register. Here, $U_z^{(2)}=W_z^{(2)}(|0\rangle\langle 0|\otimes I^{\otimes(n-1)}+|1\rangle\langle 1|\otimes V_z^{(2)})$.}
\label{NOT2}
\end{figure}
\begin{theorem}
There is a function that can be computed deterministically by an IQP circuit but cannot be by an ADIQP circuit.
\end{theorem} 
{\it Proof.}
We consider performing 1-bit classical NOT ($x\rightarrow x\oplus 1$) deterministically. In the case of IQP, it can be performed by selecting $Z_1$ as $U_z$. Next, we prove that ADIQP cannot perform such an operation deterministically. If the input register is the same as the output register, the ADIQP circuit can be written as in Fig.~\ref{NOT2}. Since the ancillary qubits are composed of $n_b$ black qubits, $n_{w,0}$ white qubits whose state is $|0\rangle^{\otimes n_{w,0}}$, and $n_{w,1}(=n-n_b-n_{w,0}-1)$ white qubits whose state is $|1\rangle^{\otimes n_{w,1}}$, the state before the $Z$-basis measurement is $[|+\rangle_oH^{\otimes (n-1)}W_z^{(2)}|+\rangle^{\otimes (n_b+n_{w,0})}|-\rangle^{\otimes n_{w,1}}+(-1)^x|-\rangle_oH^{\otimes (n-1)}W_z^{(2)}V_z^{(2)}|+\rangle^{\otimes (n_b+n_{w,0})}|-\rangle^{\otimes n_{w,1}}]/\sqrt{2}$. Here, the subscript $o$ represents the output register. Accordingly, in order to perform 1-bit NOT deterministically,
\begin{eqnarray}
|+\rangle^{\otimes (n_b+n_{w,0})}|-\rangle^{\otimes n_{w,1}}=-V_z^{(2)}|+\rangle^{\otimes (n_b+n_{w,0})}|-\rangle^{\otimes n_{w,1}}
\label{NOT}
\end{eqnarray}
is required. However, no diagonal gate in the $Z$ basis can satisfy Eq.~(\ref{NOT}). Even if the input register is different from the output register, the same argument still holds when $x=0$. \hspace{\fill}$\blacksquare$

\begin{figure*}[t]
\begin{center}
\includegraphics[width=18cm, clip]{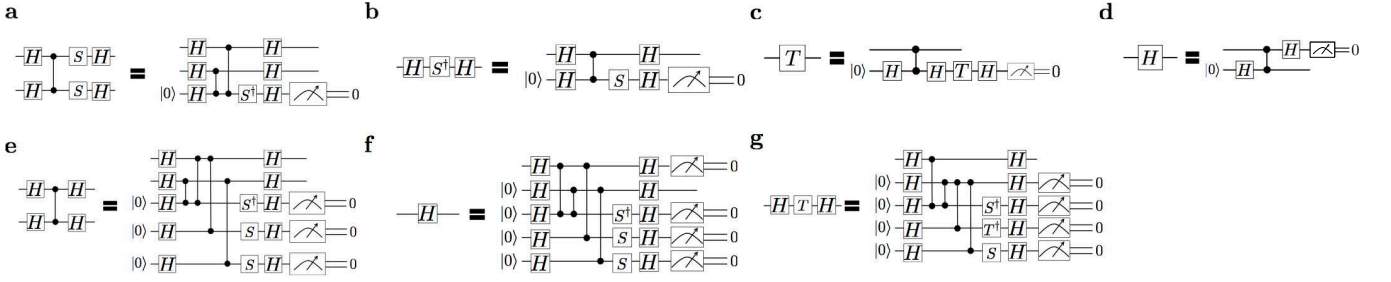}
\end{center}
\captionsetup{justification=raggedright,singlelinecheck=false}
\caption{{\bf a}-{\bf d}. The equalities used to show lemma 1. {\bf a}. Bridge operation. Here, $S\equiv Z(\pi/2)$. {\bf b}. This equality is used to remove $S$ in {\bf a}. {\bf c}. $T$ used in ADQC. {\bf d}. Hadamard gadget. {\bf e}-{\bf g}. The quantum gates on white qubits. In all cases, $0$ is output as each of the measurement outcomes with probability $1/2$. {\bf e}. $H^{\otimes 2}\Lambda(Z)H^{\otimes 2}$. Three black qubits are used as an ancillary qubit. {\bf f}. $H$. Three black qubits and one white qubit are used as ancillary qubits. The output qubit is different from the input qubit. {\bf g}. $HTH$. Three black qubits and one white qubit are used as ancillary qubits.}
\label{universal}
\end{figure*}

From the above discussions, we have shown some differences between IQP and ADIQP. However, by introducing postselection, these two models become completely equivalent from the viewpoint of complexity class. Here, we roughly define the complexity class post-A for a model A. If a decision problem is solved by a postselected A circuit in polynomial time with a probability of at least 2/3, such a problem is in the class post-A. The formal definition is introduced in~\cite{[BJS11]}. As an exception, when A is a model of universal quantum computation, we write post-A as post-BQP (bounded-error quantum polynomial time). Note that in the following lemma, $U_z$ in IQP is restricted to the diagonal gate in the $Z$ basis that is composed from a universal gate set $\{H,T,\Lambda(Z)\}$~\cite{[NC00]}.
\begin{lemma}
post-ADIQP=post-IQP
\end{lemma}
{\it Proof.} To show that an ADIQP circuit with postselection can simulate an IQP circuit with postselection, we prove that $H^{\otimes 2}\Lambda(Z)H^{\otimes 2}$, $H$, and $HTH$ can be performed on white qubits in an arbitrary order. To do so, we use the four equalities of quantum circuits shown in Figs.~\ref{universal}{\bf a}-{\bf d}. The equality shown in {\bf a} was introduced in Ref.~\cite{[FK12]} as the bridge operation. The equality shown in {\bf b} can be certified by a straightforward calculation. The equality shown in {\bf c} was introduced in Ref.~\cite{[AOKBA10]} to make ADQC. The equality shown in {\bf d} was introduced in Ref.~\cite{[BJS11]} as the Hadamard gadget. $H^{\otimes 2}\Lambda(Z)H^{\otimes 2}$ on two white qubits ({\bf e}) is performed by combining the equalities shown in {\bf a} and {\bf b}. $H$ on a white qubit ({\bf f}) is performed by combining the equalities shown in {\bf a}, {\bf b}, and {\bf d}. $HTH$ on a white qubit ({\bf g}) is performed by combing the equalities shown in {\bf a}, {\bf b}, {\bf c}, and {\bf d}. In {\bf f}, the input and output qubits are different, but these are white qubits. In {\bf f} and {\bf g}, a white qubit is used as an ancillary qubit. \hspace{\fill}$\blacksquare$

As shown in Ref.~\cite{[BMS16]}, for any degree-$3$ polynomial $f:\{0,1\}^n\rightarrow\{0,1\}$ over $\mathbb{F}_2$ satisfying $f(0^n)=0$, there is an IQP circuit $C_f$ composed of the controlled-controlled-$Z$ gate $\Lambda(\Lambda(Z))$, $\Lambda(Z)$, and $Z$, and it satisfies $\langle 0|^{\otimes n}C_f|0\rangle^{\otimes n}={\rm gap}(f)/2^n$.
\begin{lemma}
For any degree-$3$ polynomial $f:\{0,1\}^n\rightarrow\{0,1\}$ over $\mathbb{F}_2$ satisfying $f(0^n)=0$, there is an ADIQP circuit $C_f^{(2)}$ that satisfies $|\langle 0|^{\otimes (n+m)}C_f^{(2)}|0\rangle^{\otimes (n+m)}|=|{\rm gap}(f)|/2^{n+m/2}$. Here, $m=130\binom{n}{3}+3\binom{n}{2}$ and represents the number of the ancillary qubits including $25\binom{n}{3}$ white qubits. The first $n$ qubits are white qubits.
\end{lemma}
{\it Proof.} For any fixed $f$, we construct $C_f^{(2)}$ from $C_f$. To do this, we have to perform $Z$, $\Lambda(Z)$, and $\Lambda(\Lambda(Z))$ in the ADIQP circuit. Since a white qubit can be prepared as $|0\rangle$ or $|1\rangle$ as we like, $Z$ can be substantially performed on a qubit included in the first $n$ white qubits without ancillary qubits. From Fig.~\ref{universal}{\bf e}, $\Lambda(Z)$ can be performed on two arbitrary qubits included in the first $n$ qubits by using three black qubits as the ancillary qubits. As shown in chapter 4.3 in Ref.~\cite{[NC00]}, $\Lambda(\Lambda(Z))$ can be constructed from three $\Lambda(Z)$'s, four $H$'s, and three controlled-$S$ gates, and a controlled-$S$ gate can be composed from two $\Lambda(Z)$'s, four $H$'s, three $T$'s, and one $S^\dag$. In other words, $\Lambda(\Lambda(Z))$ can be constructed from nine $\Lambda(Z)$'s, 16 $H$'s, nine $T$'s, and three $S^\dag$'s. Accordingly, $\Lambda(\Lambda(Z))$ can be performed on three qubits included in the first $n$ qubits by using 25 white qubits and 105 black qubits as the ancillary qubits. From the above, at most $m$ ancillary qubits are needed to construct $C_f^{(2)}$. However, in general, some of $m$ ancillary qubits are not used. If three ancillary qubits required to perform $\Lambda(Z)$ remain, we apply $HSH$ to each of them before the $Z$-basis measurements. If 130 ancillary qubits required to perform $\Lambda(\Lambda(Z))$ remain, we apply $H^{\otimes 80}S^{\otimes 80}H^{\otimes 80}$ to 80 black qubits and $H^{\otimes 2}\Lambda(Z)H^{\otimes 2}$ to each of 25 pairs of the remaining white and black qubits before the $Z$-basis measurements. As a result, $|\langle 0|^{\otimes (n+m)}C_f^{(2)}|0\rangle^{\otimes (n+m)}|=|\langle 0|^{\otimes n}C_f|0\rangle^{\otimes n}|/\sqrt{2^m}$. \hspace{\fill}$\blacksquare$

\medskip
\begin{center}
\noindent{\bf IV. CLASSICAL SIMULATABILITY OF THE ADIQP CIRCUITS}
\end{center}
We consider the special cases of ADIQP circuits that are classically simulatable with exponentially high accuracy, i.e. in the strong sense~\cite{[NN13]}.
\begin{theorem}
Let $N_{{\rm CZ},i}$ be the number of $\Lambda(Z)$ on the $i$th black qubit.  If all $N_{{\rm CZ},i}$ are less than 1, the output probability distribution of the output register of the ADIQP circuit is classically simulatable in the strong sense.
\end{theorem} 
{\it Proof.}
If all $N_{{\rm CZ},i}$ are $0$, all of the white qubits are not connected by $\Lambda(Z)$. Accordingly, the output probability distribution of the output register can be classically simulated exactly. If all $N_{{\rm CZ},i}$ are $1$, each white qubit is connected to different black qubits. Since the output probability distribution of a 1-qubit output register of a 2-local commuting quantum computation on a product input state is classically simulatable in the strong sense~\cite{[NN13]}, the output probability distribution of the output register of the ADIQP circuit can be classically simulated in the strong sense. If all $N_{{\rm CZ},i}$ are less than $1$, for each white qubit, one of the above two observations can be applied.\hspace{\fill}$\blacksquare$

Next, using lemmas 1 and 2, we show that classical simulation of any ADIQP circuits seems to be impossible. We first consider the case of a multiplicative error.
\begin{theorem}
If the output probability distribution of any ADIQP circuit is classically simulatable up to a multiplicative error $1\le c<\sqrt{2}$, then the PH collapses at the third level.
\end{theorem}
{\it Proof.} In Ref.~\cite{[BJS11]}, it is shown that a model A that satisfies post-A=post-BQP is not classically simulatable up to a multiplicative error $1\le c<\sqrt{2}$ unless the PH collapses at the third level. From lemma 1 and post-IQP=post-BQP~\cite{[BJS11]}, post-ADIQP=post-BQP.\hspace{\fill}$\blacksquare$

Hereafter, we prove that ADIQP circuits are also hard to simulate classically up to an error $\epsilon=2^{-{\rm poly}(n)}$ in the $l_1$ norm. To do so, we use an argument similar to that used in Ref.~\cite{[BMS16]}. In the following, ${\rm BPP^{NP}}$ is a complexity class consisting of decision problems that can be solved by randomized classical polynomial-time computation given an oracle that can solve any NP problem. Moreover, ${\rm FBPP^{NP}}$ is the functional version of ${\rm BPP^{NP}}$, ${\rm P^{\#P}}$ is a complexity class consisting of decision problems that can be solved in polynomial time given an oracle that can solve any \#P problem, and ${\rm\Sigma_3P}$ is the third level of PH.

The following lemmas hold:
\begin{lemma}
Let $C_{f,x}^{(2)}$ $(x\in\{0,1\}^{n})$ be the circuit produced by applying $HZH$ to $C_f^{(2)}$ for each $k$th white qubit such that $x_k=1$ $(1\le k\le n)$. The value of $x$ is chosen uniformly at random. Assume there exists a classical polynomial-time algorithm $\mathcal{A}$ that approximates the output probability distribution of $C_f^{(2)}$ up to an error $\epsilon$ in the $l_1$ norm. Then, for any $\delta$ $(0<\delta<1)$, there is an {\rm $FBPP^{NP}$} algorithm that approximates $|\langle 0|^{\otimes (n+m)}C_{f,x}^{(2)}|0\rangle^{\otimes (n+m)}|^2$ up to an additive error $(1+o(1))\epsilon/(2^n\delta)+|\langle 0|^{\otimes (n+m)}C_{f,x}^{(2)}|0\rangle^{\otimes (n+m)}|^2/{\rm poly}(n)$ with a probability of at least $1-\delta$ over the choice of $x$.
\end{lemma}
{\it Proof.} For any $y\in\{0,1\}^n$, let $p_y={\rm Pr}[C_{f,0^n}^{(2)}\ {\rm outputs}\ y0^m]$ and $q_y={\rm Pr}[\mathcal{A}\ {\rm outputs}\ y0^m\ {\rm on\ input}\ C_{f,0^n}^{(2)}]$. From Stockmeyer's counting theorem, there is an ${\rm FBPP^{NP}}$ algorithm that produces $\tilde{q}_y$ such that $|\tilde{q}_y-q_y|\le q_y/{{\rm poly}(n)}$. Accordingly, from the triangle inequality,
\begin{eqnarray}
&&|\tilde{q}_y-p_y|\le |q_y-p_y|+\cfrac{q_y}{{\rm poly}(n)}\\
&\le&|q_y-p_y|+\cfrac{p_y+|q_y-p_y|}{{\rm poly}(n)}\\
&\le& [1+o(1)]|q_y-p_y|+\cfrac{|\langle 0|^{\otimes (n+m)}C_{f,y}^{(2)}|0\rangle^{\otimes (n+m)}|^2}{{\rm poly}(n)}.\ \ \ \
\label{tri}
\end{eqnarray}
Since $\mathcal{A}$ approximates the output probability distribution of $C_f^{(2)}$ up to an error $\epsilon$ in the $l_1$ norm, from Markov's inequality,
\begin{eqnarray}
{\rm Pr}_y[|q_y-p_y|\ge \epsilon/(2^n\delta)]\le\delta.
\label{markov}
\end{eqnarray}
From Eqs.~(\ref{tri}) and (\ref{markov}), $|\tilde{q}_y-p_y|$ is upper bounded by
\begin{eqnarray}
[1+o(1)]\cfrac{\epsilon}{2^n\delta}+\cfrac{|\langle 0|^{\otimes (n+m)}C_{f,y}^{(2)}|0\rangle^{\otimes (n+m)}|^2}{{\rm poly}(n)}
\end{eqnarray}
with probability of at least $1-\delta$.\hspace{\fill}$\blacksquare$

\begin{lemma}
${\rm Pr}_f[|\langle 0|^{\otimes (n+m)}C_f^{(2)}|0\rangle^{\otimes (n+m)}|^2\ge$$1/2^{n+m+1}]\ge 1/12$.
\end{lemma}
{\it Proof.} In Ref.~\cite{[BMS16]}, it is shown that ${\rm Pr}_f[{\rm gap}(f)^2/2^{2n}\ge 1/2^{n+1}]\ge 1/12$. From lemma 2, 
\begin{eqnarray}
&&{\rm Pr}_f[|\langle 0|^{\otimes (n+m)}C_f^{(2)}|0\rangle^{\otimes (n+m)}|^2\ge1/2^{n+m+1}]\ \ \\
&=&{\rm Pr}_f[{\rm gap}(f)^2/2^{2n+m}\ge 1/2^{n+m+1}]\\
&=&{\rm Pr}_f[{\rm gap}(f)^2/2^{2n}\ge 1/2^{n+1}]\ge 1/12.
\end{eqnarray}
\hspace{\fill}$\blacksquare$
\begin{lemma}
Let $f:\{0,1\}^n\rightarrow\{0,1\}$ be a uniformly random degree-3 polynomial over $\mathbb{F}_2$ satisfying $f(0^n)=0$. Assume that there is a classical polynomial-time algorithm that approximates the output probability distribution of any ADIQP circuit up to an error $1/(192\cdot 2^m)$ in the $l_1$ norm. Then there is an ${\rm FBPP^{NP}}$ algorithm that approximates $|\langle0|^{\otimes (n+m)}C_f^{(2)}|0\rangle^{\otimes (n+m)}|^2$ up to a multiplicative error $c=1/4+o(1)$ for at least a $1/24$ fraction of polynomials $f$.
\end{lemma}
{\it Proof.} 
From lemma 3 with $\epsilon=1/(192\cdot 2^m)$ and $\delta=1/24$, there is an ${\rm FBPP^{NP}}$ algorithm that approximates $|\langle 0|^{\otimes (n+m)}C_f^{(2)}|0\rangle^{\otimes (n+m)}|^2$ up to an additive error $(1+o(1))/2^{n+m+3}+|\langle 0|^{\otimes (n+m)}C_f^{(2)}|0\rangle^{\otimes (n+m)}|^2/{\rm poly}(n)$ with a probability of at least $23/24$. From lemma 4,
\begin{eqnarray}
&&\cfrac{1+o(1)}{2^{n+m+3}}+\cfrac{|\langle 0|^{\otimes (n+m)}C_f^{(2)}|0\rangle^{\otimes (n+m)}|^2}{{\rm poly}(n)}\\
&\le&\bigg(\cfrac{1+o(1)}{4}+\cfrac{1}{{\rm poly}(n)}\bigg)|\langle 0|^{\otimes (n+m)}C_f^{(2)}|0\rangle^{\otimes (n+m)}|^2\ \ \ \ \ \ \ \\
&=&\bigg(\cfrac{1}{4}+o(1)\bigg)|\langle 0|^{\otimes (n+m)}C_f^{(2)}|0\rangle^{\otimes (n+m)}|^2
\end{eqnarray}
with a probability of at least 1/24$(<23/288)$. \hspace{\fill}$\blacksquare$

These lemmas immediately imply the following theorem:
\begin{theorem}
Assume conjecture 1 is true. If the output probability distribution of any ADIQP circuit is classically simulatable up to an error of $1/(192\cdot 2^m)$ in the $l_1$ norm, then the PH collapses at the third level.
\end{theorem}
{\it Proof.} By combining lemma 5 with lemma 2, it is easy to obtain an ${\rm FBPP^{NP}}$ algorithm that approximates $[{\rm gap}(f)/2^n]^2$ up to a multiplicative error $1/4+o(1)$ for at least a $1/24$ fraction of polynomials $f$. Thus, by conjecture 1, it holds that ${\rm \#P\subseteq FBPP^{NP}}$ and thus ${\rm P^{\#P}\subseteq BPP^{NP}}$. Since it is known that ${\rm PH}\subseteq{\rm P^{\#P}}$~\cite{[T91],[P94]} and ${\rm BPP^{NP}}\subseteq{\rm\Sigma_3P}$, this result implies that ${\rm PH}\subseteq{\rm\Sigma_3P}$, i.e., the PH collapses at the third level. \hspace{\fill}$\blacksquare$

Finally, we consider what we should do to prove theorem 6 for $\epsilon=$const. In the ADIQP circuit, all multi-qubit gates on the concolorous qubits are prohibited. We define an ADIQP$^\ast$ circuit as a circuit that permits performing $H^{\otimes 3}\Lambda(\Lambda(Z))H^{\otimes 3}$ on any three white qubits in the ADIQP circuit.
\begin{theorem}
Assume conjecture 1 is true. If the output probability distribution of any ADIQP$^\ast$ circuit is classically simulatable up to an error of $1/192$ in the $l_1$ norm, then the PH collapses at the third level.
\end{theorem}
{\it Proof.} Unlike the argument in Ref.~\cite{[BMS16]}, in the argument to prove theorem 6, the initial state of each of the ancillary qubits is set to $|0\rangle$ to construct $C_f^{(2)}$. In other words, if we construct $C_f^{(2)}$ without setting the initial state of each of them to $|0\rangle$ when $f$ is chosen uniformly at random, we can prove theorem 6 for $\epsilon=$const. The initial state of each of the ancillary qubits that are used to perform $\Lambda(Z)$ does not have to be set to $|0\rangle$ to construct $C_f^{(2)}$. This is because, when the outputs are not $0^3$ in the quantum gate shown in Fig.~\ref{universal}{\bf e}, $Z$ is applied for each of two white qubits as byproduct operators, and these operators merely transform $C_f^{(2)}$ into $C_{f'}^{(2)}$ for some degree-$3$ polynomial $f':\{0,1\}^n\rightarrow\{0,1\}$ over $\mathbb{F}_2$ satisfying $f'(0^n)=0$. Since $Z$ can be performed without ancillary qubits as mentioned in the proof of lemma 2, if $H^{\otimes 3}\Lambda(\Lambda(Z))H^{\otimes 3}$ on three white qubits is permitted, we can prove theorem 6 for $\epsilon=1/192$ by using the same argument.\hspace{\fill}$\blacksquare$

\medskip
\begin{center}
\noindent{\bf V. CONCLUSION \& DISCUSSION}
\end{center}
In this paper, we proposed ADIQP as the subclass of commuting quantum computation suitable for many experimental settings. Although ADIQP cannot calculate 1-bit classical NOT deterministically unlike IQP, classically simulating the output probability distribution of the ADIQP circuit also seems impossible. Accordingly, ADIQP is a promising class that demonstrates quantum supremacy.

Here we discuss the verifiability of the ADIQP circuits and the difficulty in realizing them. 

The difference between IQP and ADIQP is that the graph state generated in the ADIQP circuit is only the two-colorable graph state. If the elimination of $H_i$ before the $Z$-basis measurement is allowed for any $i$ during the verification step, we can certify that the ADIQP circuit generates the correct graph state by using the stabilizer test~\cite{[HM15]}. More concretely, using the ADIQP circuit $2k+1$ times for sufficiently large $k$, we first generate $2k+1$ copies of a graph state associated with the ADIQP circuit. We then perform the stabilizer test by measuring randomly chosen $2k$ copies of them appropriately. If the test is passed, this gives a certain lower bound of the fidelity of the remaining graph state. In other words, the stabilizer test ensures that the graph state used for sampling is a desired state without measuring it. This property is very useful in experiments. Moreover, since ADIQP is a special case of IQP, the certification protocol for IQP~\cite{[HKSE16]} can also be used for ADIQP. Accordingly, ADIQP has many more verification methods than IQP.

By using the above difference, we clarify the difficulty of implementing the ADIQP circuits. In Ref.~\cite{[CL07]}, it is shown that any two-colorable graph state is equivalent to a CSS (Calderbank-Shor-Steane) state up to local unitary transformations. This equivalence implies that simulating the output probability distribution of the ADIQP circuit is as hard as simulating the probability distribution of the outcomes of the $1$-qubit measurements performed on the logical state encoded by the CSS code~\cite{[CS96],[S96]}. Accordingly, the realization of ADIQP circuits can be considered as one of the keys to realizing fault-tolerant universal quantum computation. Moreover, since the CSS code has many other applications~\cite{[TFMI16]}, such as verifiable blind quantum computing, the realization of ADIQP circuits is also meaningful for other quantum information processing schemes.

\medskip
\begin{center}
{\bf ACKNOWLEDGMENT}
\end{center}
This work was supported by the Program for Leading Graduate Schools: “Interactive Materials Science Cadet Program.”

\end{document}